\documentclass[journal]{IEEEtran}
\usepackage{graphicx}
\usepackage{amsmath}
\usepackage{amssymb}
\usepackage{amsbsy}
\usepackage{multirow}

\begin{document}
%
\title{Voice Activity Detection: Merging Source and Filter-based Information}
%

\author{Thomas Drugman, \textit{Member, IEEE}, Yannis Stylianou, \textit{Senior Member, IEEE},\\Yusuke Kida, Masami Akamine, \textit{Senior Member, IEEE}}      

\markboth{IEEE Signal Processing Letters}%
{Shell \MakeLowercase{\textit{et al.}}: Bare Demo of IEEEtran.cls for Journals}

\maketitle

\begin{abstract}
Voice Activity Detection (VAD) refers to the problem of distinguishing speech segments from background noise. Numerous approaches have been proposed for this purpose. Some are based on features derived from the power spectral density, others exploit the periodicity of the signal. The goal of this paper is to investigate the joint use of source and filter-based features. Interestingly, a mutual information-based assessment shows superior discrimination power for the source-related features, especially the proposed ones. The features are further the input of an artificial neural network-based classifier trained on a multi-condition database. Two strategies are proposed to merge source and filter information: feature and decision fusion. Our experiments indicate an absolute reduction of 3\% of the equal error rate when using decision fusion. The final proposed system is compared to four state-of-the-art methods on 150 minutes of data recorded in real environments. Thanks to the robustness of its source-related features, its multi-condition training and its efficient information fusion, the proposed system yields over the best state-of-the-art VAD a substantial increase of accuracy across all conditions (24\% absolute on average).
\end{abstract}


%


\let\thefootnote\relax\footnotetext{
\\Authors are with Toshiba Cambridge Research Laboratory. \textit{Address: 208 Cambridge Science Park, Milton Road, Cambridge, CB4 0GZ, UK},\emph{Phone: +44 1223 436900}, \emph{Email:} yannis.stylianou@crl.toshiba.co.uk.}

\section{Introduction}
\label{sec:Intro}

Voice Activity Detection (VAD) refers to the problem of distinguishing speech segments from background noise in an audio stream. This is a fundamental task which finds a wide range of applications in voice technology: speech coding \cite{G729}, automatic speech recognition (ASR, \cite{ASR}), audio surveillance and monitoring, speech enhancement, or speaker and language identification \cite{SpeakerID}. In the workflow of these applications, VAD is generally involved as the very first block. As a consequence, the main characteristics expected from a VAD algorithms are generally a high efficiency and robustness to noise, as well as a low computational latency.

Numerous studies have addressed the problem of VAD in the literature. Generally speaking, a VAD method consists of two successive steps: feature extraction and a discrimination model. Early works focused on energy-based features, possibly combined with the zero-crossing rate (ZCR) \cite{Lamel, Kotnik}. These features are however highly affected in the presence of additive noise. Therefore, various other features have been proposed: autocorrelation-based features \cite{Kingsbury, Kristjansson, Hansen}, Mel-Frequency Cepstral Coefficients (MFCCs) \cite{Kristjansson}, line spectral frequencies \cite{Marzinzik}, a cepstral distance \cite{Haigh}, the skewness and kurtosis of the linear prediction (LP) residual \cite{Nemer} or periodicity-based features \cite{Tucker, Ishizuka, Hansen}. Some other methods are based on a statistical model of the Discrete Fourier Transform (DFT) coefficients \cite{Sohn, Ramirez}. Other approaches exploit the fact that the speech and noise signals should have different variability properties \cite{Ghosh, Ramirez2}. Finally, some studies have addressed the use of a combination of multiple features. These works differ by the way the features are combined: using a linear combination where the weights are trained via a minimum classification error in \cite{Kida}, a linear or a kernel discriminant analysis in \cite{Soleimani} or a principal component analysis in \cite{Hansen}.

The resulting acoustic information is then generally the input of a statistical model whose goal is to draw a decision about the presence or not of speech. Proposed approaches differ in whether they use a supervised framework or not. In the former case, several models have been used: Gaussian Mixture Model (GMM, \cite{GMM}), Hidden Markov Model (HMM, \cite{HMM}) or Multi-Layer Perceptron (MLP, \cite{GMM}). Some other works state that the drawbacks of a supervised method are that large amounts of labeled training data are required and that they are sensitive to a mismatch between training and testing conditions \cite{Hansen, Germain}. As a consequence, unsupervised approaches have been recently proposed in \cite{Ying, Hansen, Germain}.

With respect to the state of the art, the main contributions of this paper are the following: \emph{i)} to propose the use of robust source-related features for VAD purpose, \emph{ii)} to assess the relative performance of source and filter-based features, \emph{iii)} to investigate the best strategies to merge information from various feature sets, \emph{iv)} to compare the proposed VAD system with existing algorithms on \emph{real data}, \emph{v)} to examine the generalization capabilities of a supervised approach when trained on a multi-condition dataset. Note that the two last points must be moderated as recent studies conducted VAD experiments on real-life videos \cite{Misra, Eyben}, possibly with a multi-condition training approach \cite{Eyben}.



The paper is organized as follows. The proposed method is described in Section \ref{sec:Proposed}. The protocol used throughout our experiments is presented in Section \ref{sec:Protocol} and the results are discussed in Section \ref{sec:Results}. Section \ref{sec:Conclu} finally concludes the paper.

\section{Proposed Technique}
\label{sec:Proposed}

According to the mechanism of voice production, speech is considered as the result of a glottal flow (also called \emph{source} or \emph{excitation} signal) filtered by the vocal tract cavities \cite{CSLReview}. This physiological process motivates the goal of this paper as we believe it to be essential that a VAD exploits information from both these two complementary components of speech. The proposed VAD approach will be shown in Section \ref{sec:Results} to carry out a significant improvement over the best state-out-the-art approach. It is worth emphasizing on that this would not be possible without the \emph{combined} effect of 4 main factors: the joint use of filter and source-related information, the design of robust source features, an efficient strategy of information fusion and a multi-condition training.

\subsection{Filter-based Features}
\label{ssec:SpectralEnvelope}

According to the source-filter model of speech, the \emph{spectral envelope}, defined as a smooth function passing through the prominent peaks of the spectrum \cite{SpectralEnvelope}, is the transfer function of the filter. Various ways to parameterize the spectral envelope have been proposed in the literature. In this work, the following representations are considered: the Mel Frequency Cepstral Coefficients (MFCCs, \cite{MFCC}), the Perceptual Linear Prediction coefficients (PLP, \cite{PLP}) and the Chirp Group Delay (CGD) of the zero-phase signal which is a robust high-resolved representation of the filter resonances \cite{CGD}. A vector of 13 coefficients is used for each feature type. The advantage to use such parameters is that they have been already shown to be efficient for ASR or speaker recognition purpose \cite{CGD, SpeakerID2}, and can generally be of high interest in any speech technology application following the VAD. Most of the time, their computation is therefore already required and their integration in a VAD system can therefore be achieved at a very low computational cost.

\subsection{Source-related Features}
\label{ssec:Excitation}

The glottal flow has been recently shown to be useful in various voice technologies \cite{DrugmanThesis, CSLReview}. Despite promising advances, it has been acknowledged in \cite{CSLReview} that the weakest point of current glottal source processing algorithms is clearly related to their lack of robustness. It is therefore a challenging and still open problem to design source-related features for applications in adverse environments. One issue is the strong degradation of glottal flow estimation techniques when the speech signal gets noisier \cite{GFEstimation}. When working in adverse conditions, it is consequently preferable to use indirect measurements derived either from the speech signal or from the LP residue. In this work, we aim at using robust source-related features which are compatible with the noisy environments targeted by our VAD.

Various existing studies have already used excitation information for VAD. The periodicity of the speech signal has been exploited in \cite{Kingsbury, Kristjansson, Tucker, Ishizuka, Hansen}. Furthermore, features extracted from the LP residual have been used in \cite{Nemer}. In this work, we consider some of these features already proposed for VAD purpose, as well as some new other source-related measurements. Two popular features used from the early attempts are the log-energy of the speech signal \cite{Lamel} and the zero-crossing rate (ZCR, \cite{Kotnik}). In \cite{Nemer}, Nemer et al. proposed the use of high-order statistics of the LP residual. As suggested in that study, we included the skewness and kurtosis of the LP residual which are known to respectively characterize the polarity of the speech signal \cite{Polarity} and the sparsity of the excitation \cite{SparseLP} at the glottal closure instants. Sadjadi recently proposed in \cite{Hansen} a VAD system using 4 voicing measures: the so-called harmonicity and clarity features derived from the average magnitude difference function (AMDF), the normalized LP error \cite{Nemer} and the Harmonic Product Spectrum (HPS, \cite{Hansen}).



In addition to the aforementioned features, we include three other source-related measurements. These latter features were proposed in previous studies and were here selected for their robustness properties. The first is the Cepstral Peak Prominence (CPP) which was originally proposed in \cite{Hillenbrand} for the prediction of breathiness ratings. CPP is a measure of the amplitude of the cepstral peak at the hypothesized fundamental period. The two other features are extracted from the Summation of the Residual Harmonics (SRH) algorithm \cite{SRH}, a robust pitch tracker. The SRH criterion quantifies the level of voicing by taking into account the harmonics and inter-harmonics of the residual spectrum. The two features used in this work, referred to as $SRH$ and $SRH^*$ differ by the energy normalization or not of the residual spectrum. Note that the implementations of CPP and SRH are available from the COVAREP project \cite{COVAREP}.

\subsection{ANN-based Classification and Information Fusion}
\label{ssec:ANN}

For our classification experiments, we opted for an ANN for its discriminant properties, its ability to model non-linear relations and for the convenience of the posterior probabilities it generates. Each ANN is made of a single hidden layer consisting of neurons whose activation function is an hyperbolic tangent sigmoid transfer function. As any parameter used by the proposed technique, the number of neurons was set on the development data. Performance was very similar using between 32 and 128 neurons, and we fixed this parameter to 32 in the remainder of this paper. The output layer is a simple neuron with a sigmoid function suited for a binary VAD decision. Note that we also tried to make use of recurrent neural networks. This however did not lead to a particular gain in performance while it increased the computational load.

Before being fed to the ANN, the feature vector $x_t$ at time $t$ goes through two processing steps. First, the feature trajectories are smoothed using a median filter with a width of 11 frames (5 on each side). Working with a frame shift of 10 ms, this roughly corresponds to the phone scale. This operation allows to remove possible spurious values. Secondly, contextual information is added by including the first and second derivatives, computed using the following finite difference equation: $x'_t=\frac{1}{N}\sum_{i=1}^{N}{x_{t}-x_{t-i}}$. To keep working at the phone level, the number $N$ of contextual frames is set to 10. When in test, the ANN outputs the posterior of speech activity. As a last post-process, the posterior trajectories are smoothed out by a median filter whose width is again set to 11 frames so as to remove possible erroneous isolated decisions.

Our goal being to combine various sets of features, we consider two strategies to merge their information: feature fusion and decision fusion (also called early and late fusion). In the feature fusion case, synchronous feature vectors are simply concatenated and a single ANN is trained. In the decision fusion case, one specific ANN is trained for each feature set. Each ANN outputs a trajectory of posteriors, and the trajectories from the various ANNs are further merged to derive one final posterior value. Several strategies to combine the posteriors have been proposed in \cite{Kittler}. In this work, we have tried the arithmetic and the geometrical mean (corresponding to the sum and product rule in \cite{Kittler}). The differences in performance that we noticed were however negligible, and the geometrical mean is used throughout the rest of this paper.

\section{Experimental Protocol}
\label{sec:Protocol}

\subsection{Speech databases}
\label{ssec:Databases}
For the training of the proposed technique, our goal was to use a corpus containing a large diversity of speakers and noisy conditions. We chose a subset of 1500 files from the TIMIT database \cite{TIMIT} from 300 speakers. As the original utterances were recorded in clean studio conditions, the advantage of this approach is that the labels can be easily obtained by using a simple energy threshold to extract the speech endpoints. For each file, noise was then artificially added at two SNR levels: 0 and 10 dB, leading to a total of 3000 files. For each file, the noise was randomly selected among 4 types from the Noisex-92 database: babble, car, factory and jet noises. Note that we added 2 seconds of noise before and after each utterance so that the database is roughly balanced between speech activity and background noise. We expect that this multi-condition training set is sufficiently diversified for the classifier to be effective in various (possibly unseen) environments and with new speakers. The development set consists of a 5\% held-out portion of the training set.

The testing corpus is a manually annotated proprietary database containing real data recorded in 5 places: mall, kitchen, street, station and living room. Various sources of noise are therefore covered and encompass TV in the background, people talking nearby, cooking, cars passing by, etc. The data consists of Japanese read speech from 5 speakers using either a tablet or a smartphone. The main characteristics of the testing database are summarized in Table \ref{tab:TestingDatabase}. Note that the averaged SNR only reflects one aspect of the noise, and that other characteristics such as its dynamics and its spectral shape might be a preponderant source of performance degradation.

\begin{table}[!ht]
\caption{Characteristics of the testing database.}
\centering
\begin{tabular}{| c | c | c | c | c | c || c |}
\hline    
\textbf{Environ.} & Kitchen & Mall & Station & Living & Street & Overall\\
\hline    
\textbf{Dur.} (min) & 49 & 19.8 & 20 & 42.3 & 20.5 & 151.6\\
\hline
\textbf{Av.SNR} (dB) & 7.3 & 6.9 & 13.5 & 18.2 & 15.1 & 12.2\\
\hline
\textbf{\% of speech} & 12.3 & 20.2 & 20.5 & 22.6 & 18.9 & 18.9\\
\hline
\end{tabular}
\label{tab:TestingDatabase}
\vspace{-16pt}
\end{table}

\subsection{Assessment Metrics}
\label{ssec:Metrics}

As a first metric to quantify the discrimination power of each feature individually, we use the normalized mutual information \cite{DrugmanFS}, defined as the mutual information (MI) of the feature with the class labels divided by the class entropy. The normalization ensures an intuitive interpretation with values ranging between 0 and 1. This measure has also the advantage to be independent from the subsequent classifier. The computation of mutual information is here carried out via a histogram approach \cite{DrugmanFS}. The number of bins is set to 50 for each feature dimension, which results in a trade-off between an adequately high number for an accurate estimation, while keeping sufficient samples per bin. Class labels correspond to the presence (or not) of speech. 


To assess the perfomance after classification, two metrics are used. These two measures respectively characterize the frame and the utterance levels. By varying a decision threshold $\theta$, a Receiving Operating Characteristics (ROC) curve can be obtained. The first metric is the so-called \emph{equal error rate} (EER), which corresponds to the location on a ROC curve where the false accept rate and false reject rate are equal. The second metric quantifies the ability to detect the endpoints of speech utterances. For this purpose, we use the F1 score (maximized over $\theta$ in the dev set) as a single measure combining both precision and recall. The F1 score ranges from 0 to 1, where 1 implies a perfect classification. The correctness of a speech segment with regard to a reference is conform to the CENSREC-1-C criteria defined in \cite{CENSREC}. Note that before being assessed at the utterance level, the vector of binary decisions goes through an hangover scheme \cite{Hangover} consisting of a morphological closing (i.e. a dilatation followed by an erosion) with a time constant of 600 ms and a length extension of 200 ms on each side. Note that the same hangover scheme was applied to all techniques for the computation of the utterance-level results.

\subsection{Comparison with state-of-the-art techniques}
\label{ssec:Comparison}
Four state-of-the-art VAD systems are used for comparison purpose: the G.729B algorithm \cite{G729}, Shon's statistical model-based VAD \cite{Sohn}, Ying's unsupervised technique based on sequential Gaussian mixture models \cite{Ying} (whose code was kindly shared by Dongwen Ying), and Ghosh's VAD using long-term signal variability \cite{Ghosh}. As for the proposed technique, each of these methods makes use of a decision parameter which was tuned to optimize the EER and F1 scores, as discussed in Section \ref{ssec:Metrics}.

\section{Results}
\label{sec:Results}

\subsection{Mutual information-based assessment}
\label{ssec:MI}

The results of the MI-based assessment are presented in Table \ref{tab:MI}. For the filter-based features (MFCC, CGD and PLP), MI values have been averaged across the 13 coefficients. Note also that, for each feature, these results are averaged across static, first and second derivatives values. It can be seen that CGD gives the best results among the spectral envelope representations. Among the source-related features, the three proposed features interestingly provide the best results. They are followed by 3 features used in \cite{Hansen}: HPS, harmonicity and clarity. This latter feature achieves a MI value comparable to that of the LP kurtosis and of the log-energy. As mentioned in Section \ref{ssec:Excitation}, designing robust source-related features is a challenging problem. The fact that the 3 proposed features yield better performance can be explained as follows: \emph{i)} time-domain features are expected to be more sensitive to noise and working either in the spectral or cepstral domain turns out to be more appropriate, \emph{ii)} SRH features outperform HPS because they exploit interharmonics as well as the LP residue which allows to minimize the effects of both the vocal tract resonances and of the noise \cite{SRH}.

\begin{table}[!ht]
\caption{Mutual information-based feature assessment}
\centering
\begin{tabular}{| c | c | c | c | c | c | c | c |}
\hline
\textbf{Feature} & MFCC & CGD & PLP & Energy & ZCR & Kurt. & Skew.\\
\hline
\textbf{MI} (\%) & 14.3 & 17.2 & 13.6 & 27.3 & 22.6 & 27.3 & 19.0\\
\hline
\hline
\textbf{Feature} & Harm. & Clar. & LP err. & HPS & CPP & SRH & SRH*\\
\hline
\textbf{MI} (\%) & 29.7 & 27.2 & 17.1 & 32.9 & 36.3 & 38.7 & 51.8\\
\hline
\end{tabular}
\label{tab:MI}
\vspace{-16pt}
\end{table}


\subsection{Classification results}
\label{ssec:Classification}

For these experiments, we consider various sets of features: 13 MFCCs, 13 CGDs, 13 PLPs, the 4 voicing features (Harmonicity, Clarity, LP error and HPS) used in Sadjadi's paper \cite{Hansen}, and the 3 new source-based features (CPP, SRH and SRH*) which have not been used for VAD purpose yet. The two last sets of features will be referred to as \emph{Sadjadi} and \emph{New} in the following. The performance of these 5 feature sets is shown in Table \ref{tab:Results1}. Two main conclusions, which corroborate our observations from Section \ref{ssec:MI}, can be drawn from these results: \emph{i)} for VAD purpose, source-related features are more relevant than those characterizing the filter. Among them, the \emph{Sadjadi} and \emph{New} feature sets achieve similar performance; \emph{ii)} across the filter representations, the CGD features, whose robustness was already highlighted in \cite{CGD} for ASR purpose, turn out to be the most efficient. Nonetheless, since MFCCs are widely used in various speech technology applications and that their extraction is likely to be required anyways, we chose to use them as filter-based features in the rest of this paper. 

\begin{table}[!ht]
\caption{Classification results using the 5 feature sets.}
\centering
\begin{tabular}{| c | c | c | c | c | c |}
\hline
\textbf{Feature Set} & MFCC & CGD & PLP & Sadjadi & New\\
\hline
\textbf{1-EER} (in \%) & 87.9 & 90.2 & 87.6 & 93.7 & 94.0\\
\hline
\textbf{F1 score} (in \%) & 77.1 & 79.2 & 75.5 & 86.8 & 86.7\\
\hline
\end{tabular}
\label{tab:Results1}
\vspace{-8pt}
\end{table}

In the second part of our experiments, we investigated the combination of different feature sets either at the feature or the decision level (see Section \ref{ssec:ANN}). The results are displayed in Table \ref{tab:Results2}, where N and S respectively stand for the \emph{New} and \emph{Sadjadi} feature sets. Note that Table \ref{tab:Results2} only shows the EER-based results; similar conclusions could be however drawn from the F1 scores. Interestingly, it can be observed that in all cases the decision fusion scheme outperforms feature fusion, by 3\% in absolute on average. Feature fusion even led to a degradation in 3 out of the 4 cases. This is important because feature concatenation is conventionally used in most existing approaches. One possible reason to explain this is the curse of dimensionality \cite{Bellman}: as the dimensionality of the feature vector increases, it becomes more and more difficult to accurately model the data, as an ever increasing number of samples is required. Although the association of the two excitation-based feature sets (S+N) yields already a high performance, the best results are obtained when they are combined with MFCCs. This is however only true when using the decision fusion. In the rest of our experiments, the system based on these 3 feature sets and using decision fusion will be referred to as the proposed VAD system.

\begin{table}[!ht]
\caption{Classification results (1-EER, in \%) using a combination of feature sets.}
\centering
\begin{tabular}{| c | c | c | c | c |}
\hline
\textbf{Combination} & MFCC+S & MFCC+N & S+N & MFCC+S+N\\
\hline
\textbf{Feature fusion} & 93.6 &  89.8 & 94.9 & 90.9\\
\hline
\textbf{Decision fusion} & 94.8 & 95.4 & 95.3 & 95.8\\
\hline
\end{tabular}
\label{tab:Results2}
\vspace{-16pt}
\end{table}

The comparative evaluation with state-of-the-art techniques is summarized in Table \ref{tab:Results3} for the 5 different environments and using the F1 score. Note that all the observations that will be made hereafter were also corroborated using the EER metric. Three main conclusions can be drawn from Table \ref{tab:Results3}. First, it can be noticed that across all conditions the proposed system clearly outperforms existing methods, sometimes by a large increase of the F1 score. This is especially true in the kitchen, living room and mall environments, where existing algorithms tend to fail dramatically. This is mostly due to the fact that the corresponding recordings contain sporadic impulsive noises such as cough, laughter or cooking, whose dynamics can sometimes be similar to that of speech. These environments are therefore much more challenging than the street and station conditions which are rather stationary. Secondly, it is worth reminding that the four state-of-the-art techniques used in this comparison are based on the power spectral density, and therefore discard any source-related information. This further supports our results from Tables \ref{tab:MI} and \ref{tab:Results1} that excitation-based features are necessary in an efficient VAD system. Finally, despite the mismatch between training and testing data, the proposed algorithm works well in all environments. This makes us think that the generalization capabilities of the proposed system are high, and that it can potentially adapt to any new environment, speaker, language or sensor. This is likely due to the robustness of the source-related features as well as the ability of the ANN to capture the speech patterns through the multi-condition training.

\begin{table}[!ht]
\caption{Comparison with state-of-the-art methods (F1 scores, in \%).}
\centering
\begin{tabular}{| c | c | c | c | c | c |}
\hline
 & G.729B \cite{G729} & Sohn \cite{Sohn} & Ying \cite{Ying} & Ghosh \cite{Ghosh} & Prop.\\
\hline
\textbf{Kitchen} & 37.8 & 43.9 & 51.2 & 44.6 & \textbf{89.2}\\
\hline
\textbf{Mall} & 33.2 & 69.6 & 70.9 & 67.0 & \textbf{89.4}\\
\hline
\textbf{Station} & 67.6 & 85.5 & 84.2 & 93.6 & \textbf{95.6}\\
\hline
\textbf{Living} & 28.0 & 46.9 & 47.8 & 45.1 & \textbf{93.3}\\
\hline
\textbf{Street} & 72.1 & 84.5 & 81.8 & 94.9 & \textbf{97.7}\\
\hline
\hline
\textbf{Average} & 47.7 & 66.1 & 67.2 & 69.1 & \textbf{93.0}\\
\hline
\end{tabular}
\label{tab:Results3}
\vspace{-16pt}
\end{table}


\section{Conclusion}
\label{sec:Conclu}

The goal of this paper was to investigate the joint use of source and filter-based features for VAD purpose. The main conclusions of this study are the following: \emph{i)} source-related features, and especially the 3 proposed features, have a better discrimination power and their use in an efficient VAD system is necessary, \emph{ii)} as a strategy to merge different sources of information, decision fusion outperforms feature fusion, \emph{iii)} the resulting proposed system, combining source and filter-based information, gives a significantly better performance compared to state-of-the-art methods, \emph{iv)} the robustness of source-related features combined with the generalization capabilities of neural networks makes the proposed approach perform very well in unseen conditions. Features used in this paper can be extracted with the following toolkit: \emph{tcts.fpms.ac.be/$\sim$drugman/files/VAD.zip}.



\ifCLASSOPTIONcaptionsoff
  \newpage
\fi

\end{document}